\documentclass[conference,a4paper]{IEEEtran}
\IEEEoverridecommandlockouts
\usepackage{cite}
\usepackage{amsmath,amssymb,amsfonts}
\usepackage{algorithmic}
\usepackage{graphicx}
\usepackage{textcomp}
\usepackage{xcolor}

\usepackage{multirow}
\usepackage{graphicx}
\usepackage{subcaption}
\usepackage{amsmath}
\usepackage{tabularx}
\usepackage{filecontents}
\usepackage{multirow}
\usepackage{booktabs}  
\usepackage{threeparttable}
\usepackage{colortbl}
\usepackage{siunitx}
\usepackage{tikz}

\newcommand{\Name}{Merlin}

\def\BibTeX{{\rm B\kern-.05em{\sc i\kern-.025em b}\kern-.08em
    T\kern-.1667em\lower.7ex\hbox{E}\kern-.125emX}}
\begin{document}

\title{Offline to Online Learning for Real-Time Bandwidth Estimation}

\author{}
\author{\IEEEauthorblockN{Aashish Gottipati\textsuperscript{*}, Sami Khairy\textsuperscript{\textdagger}, Gabriel Mittag\textsuperscript{\textdagger}, Vishak Gopal\textsuperscript{\textdagger}, and Ross Cutler\textsuperscript{\textdagger}}
\IEEEauthorblockA{\textsuperscript{*}Department of Computer Science, University of Texas at Austin, Austin, Texas \\ \textsuperscript{\textdagger}Microsoft, Redmond, Washington}}

\newcommand\copyrighttext{%
\footnotesize \textcopyright 2025 IEEE. Personal use of this material is permitted.
Permission from IEEE must be obtained for all other uses, in any current or future
media, including reprinting/republishing this material for advertising or promotional
purposes, creating new collective works, for resale or redistribution to servers or
lists, or reuse of any copyrighted component of this work in other works.}

\newcommand\copyrightnotice{%
\begin{tikzpicture}[remember picture,overlay]
\node[anchor=south,yshift=10pt] at (current page.south)
{\fbox{\parbox{\dimexpr\textwidth-\fboxsep-\fboxrule\relax}{\copyrighttext}}};
\end{tikzpicture}%
}

\maketitle
\copyrightnotice

\begin{abstract}
Real-time video applications require accurate bandwidth estimation (BWE) to maintain user experience across varying network conditions. However, increasing network heterogeneity challenges general-purpose BWE algorithms, necessitating solutions that adapt to end-user environments. While widely adopted, heuristic-based methods are difficult to individualize without extensive domain expertise. Conversely, online reinforcement learning (RL) offers ease of customization but neglects prior domain expertise and suffers from sample inefficiency. Thus, we present \Name{}, an imitation learning-based solution that replaces the manual parameter tuning of heuristic-based methods with data-driven updates to streamline end-user personalization. Our key insight is that \textit{transforming} heuristic-based BWE algorithms into neural networks facilitates data-driven personalization. \Name{} utilizes Behavioral Cloning to efficiently learn from offline telemetry logs, capturing heuristic policies without live network interactions. The cloned policy can then be seamlessly tailored to end user network conditions through online finetuning. In real intercontinental videoconferencing calls, \Name{} matches our heuristic's policy with no statistically significant differences in user quality of experience (QoE). Finetuning \Name{}'s control policy to end-user environments enables QoE improvements of up to $7.8\%$ compared to the heuristic policy. Lastly, our IL-based design performs competitively with current state-of-the-art online RL techniques but converges with $80\%$ fewer videoconferencing samples, facilitating \textit{practical} end-user personalization.
\end{abstract}


\section{Introduction}
Real time video applications are pervasive and rely on accurate bandwidth estimation (BWE) to adjust video encoding bitrates without exhausting network resources. BWE addresses the challenge of estimating the bottleneck link-- the network link with the least available bandwidth, which dictates the rate of information flow across the network. Despite the existence of multiple general-purpose BWE algorithms~\cite{fang_reinforcement_2019, bob, hrcc, gemini}, increasing network heterogeneity continues to challenge these methods. Nearly $80\%$ of videoconferencing users have reporting degraded quality of experience (QoE) on popular platforms such as Zoom, Teams, and Skype~\cite{matulin2021user}. To address the limitations of existing BWE algorithms in heterogeneous environments, researchers have attempted to tailor these methods to end-user environments~\cite{naseer2022configanator}. However, personalizing existing algorithms to individual contexts in a \textit{practical} manner remains a challenge.

For example, Google's Congestion Control (GCC) in WebRTC~\cite{bergkvist2012webrtc} employs statistical methods like trendline filters and unscented Kalman filters to adapt capacity estimates by modeling networks as nonlinear systems. While WebRTC is the most widely adopted real-time communication (RTC) stack, tailoring GCC to specific end-user environments is extremely challenging. Personalizing GCC requires manually tuning numerous parameters, for the Kalman filter (state noise variance, exponential moving average factor, initial system error variance) and the rate-controllers (delay threshold adjustment rates, send rate modification factors, loss factor, pacing factor, estimated maximum bitrate threshold). These parameters were carefully calibrated by domain experts through extensive experimentation~\cite{bergkvist2012webrtc}, a process impractical to replicate for millions of RTC users. Moreover, as network complexity increases, determining optimal parameters becomes increasingly difficult, even for domain experts~\cite{tcp-conf}, highlighting the limitations of relying solely on human intuition.

In contrast, deep reinforcement learning (RL) models have demonstrated impressive ability for real-time control under complex domains, while enabling ease of updates through new data observations and tuned reward functions~\cite{rl-finetune}. Although these properties are desirable, RL agents are conventionally trained from a blank slate in online environments, neglecting the vast amount of prior domain knowledge. This trial and error strategy requires a significant number of training samples to converge to an acceptable policy~\cite{schulman_proximal_2017}. In the case of videoconferencing, the sample complexity is on the order of hundreds of thousands of live videoconferencing calls, making them impractical to adopt for continuous end-user personalization~\cite{fang_reinforcement_2019}.

Our key insight to address the \textit{practical} limitations of end-user personalization is to \textit{transform} existing heuristic-based BWE algorithms into neural networks with imitation learning (IL)~\cite{imitation-learning} and to finetune the neural representations on end-user environments. We leverage offline telemetry logs from heuristic-based BWE algorithms as demonstrations, combining them with IL and finetuning techniques to derive more specialized BWE policies. Rather than carefully calibrating an existing heuristic or training a new BWE from scratch, we instead choose the middle ground and blend the two approaches. Combining heuristics with data-driven methods offers two key advantages: (1) algorithm parameters can be readily tailored according to QoE objectives through gradient-based updates; (2) the vast amounts of domain expertise encoded in heuristic-based methods can greatly reduce the large sample complexity associated with trial-and-error from scratch.

However, \textit{transforming} existing heuristic-based BWE algorithms into neural networks presents many challenges. First, deriving a policy from a limited set of heuristic demonstrations may not capture the true, generalizable heuristic policy, potentially leading to poor estimates under unfamiliar conditions and degrading user QoE~\cite{sage}. Second, handcrafted heuristic features may not directly translate to data-driven methods. Inadequate state representation can hinder the estimator's ability to capture the relationship between network states and heuristic actions~\cite{finn2016guided}.

Consequently, we propose \Name{}, an IL-based solution that replaces the manual parameter tuning of heuristics with data-driven updates to streamline end-user personalization. \Name{} is trained to imitate an Unscented Kalman Filter (UKF), a hand-crafted estimator previously deployed in production on Microsoft Teams. \Name{} learns strictly from offline telemetry logs via behavioral cloning (BC), an IL method that utilizes supervised learning~\cite{bc-from-obs}. \Name{}'s hybrid design facilitates data-driven personalzation without sacrificing the domain knowledge of heuristic-based methods. We demonstrate that \Name{}'s learned policy generalizes from offline experience to the real world on live, intercontinental videoconferencing calls, where \Name{} matches UKF's policy with no statistically significant movements in terms of objective QoE metrics. Furthermore, we show that specializing \Name{}'s control policy is possible through a small number of online data-driven parameter updates, improving QoE by up to $7.8\%$ compared to our heuristic. Lastly, we show that our method is significantly more sample efficient compared to existing online RL methods, consistently converging to high quality policies with $80\%$ less training samples.

In summary, our contributions are as follows:
\begin{enumerate}
\item We introduce \Name{}, a novel approach to BWE that \textit{transforms} existing heuristic-based BWE algorithms into neural networks using IL, combining domain expertise with data-driven methods to enable practical end-user personalization.
\item We demonstrate that \Name{} can effectively learn and generalize heuristic policies from offline telemetry logs to real-world, intercontinental videoconferencing calls, matching heuristic performance with no statistically significant deviations.
\item We show that \Name{} enables efficient personalization through a small number of online data-driven updates, improving QoE by up to $7.8\%$ compared to the heuristic estimator while being significantly more sample-efficient than existing online RL methods.
\end{enumerate}

\section{Related Work}

\begin{figure}[t!]
    \includegraphics[width=0.95\columnwidth, trim={0 14cm 0 6.75cm}, clip]{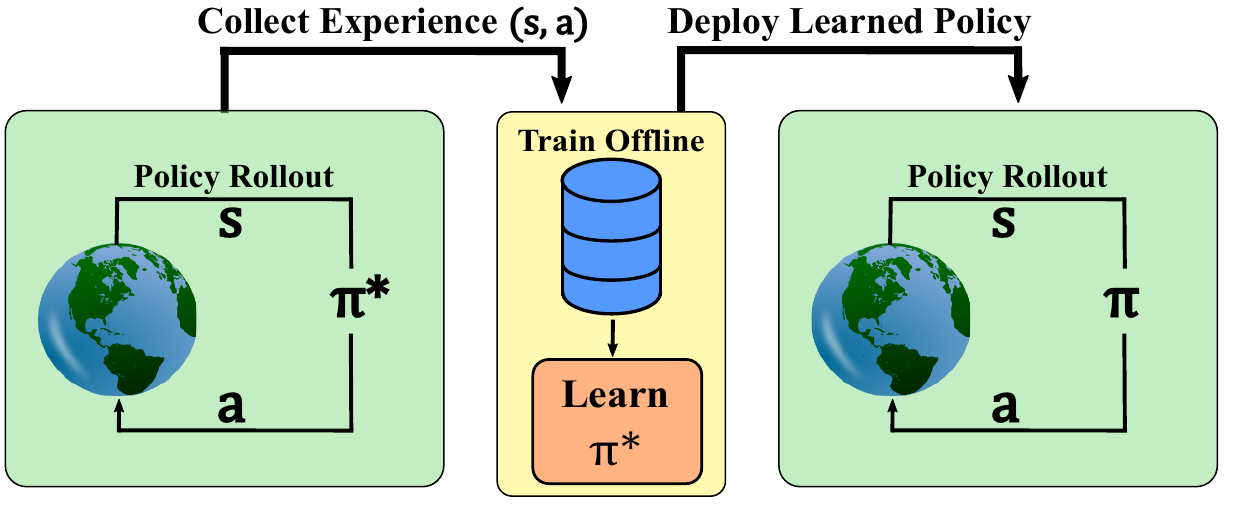}
    \vspace{-3mm}
	\caption{Learning from Offline Demonstrations.}
	\label{fig:imitation_procedure}
    \vspace{-6mm}
\end{figure}

\label{sec:related}

Recent works have sought to leverage heuristic algorithms for model training. Eagle~\cite{emara2020eagle} adopts an heuristic-oriented approach, seeking to match BBR~\cite{bbr} via online RL. DuGu~\cite{dugu} utilizes an online IL approach to mimic a custom congestion control (CC) oracle for short-term video uploads. Zhou et al. propose Concerto~\cite{concerto}, a BC-based method that leverages oracle estimates to select the best video bitrate from a discrete set of encodings. Indigo~\cite{yan_pantheon:_2018} learns to adjust the number of allowable in-flight packets from a set of ideal congestion control oracles; however, as opposed to prior work, \Name{} learns to estimate the available network resources directly with offline telemetry logs collected from heuristic-based methods. 

Later works such as Gemini~\cite{gemini}, HRCC~\cite{hrcc}, HybridRTS~\cite{hybridrts}, BoB~\cite{bob}, SAFR~\cite{safr}, OnRl~\cite{onrl}, and Libra~\cite{libra} build non-standalone estimators directly on top of heuristic methods or utilize heuristics as fall-back mechanisms to mitigate tail-end predictions of learned agents. Along similar lines, Zhang et al. explore fusing cloned heuristic features with online RL-based models for video bitrate prediction~\cite{loki}. In contrast to these works, we propose \Name{} a standalone, data-driven approach to BWE that does not rely on auxiliary estimators at runtime.

Lastly, the most similar work to ours is Sage~\cite{sage}. Sage builds upon the vast efforts of heuristic-based methods to learn 
a data-driven CC algorithm via offline RL. In contrast to Sage, 
we focus on \textit{transforming} heuristic-based BWE algorithms
into neural networks with offline IL to facilitate \textit{practical} personalization. We demonstrate the potential for specializing heuristic policies through online finetuning as opposed to stitching together a policy from a mixture of past offline experiences. 

\section{Merlin}
\label{sec:methods}

{\bf Overview.} Traditional heuristic-based bandwidth estimators leverage extensive domain knowledge to produce robust bandwidth estimates; however, these methods are often challenging to specialize. Conversely, emerging RL-based estimators offer enhanced configurability but neglect prior domain expertise and exhibit high sample complexity. To address these limitations, we propose an approach that combines offline IL with finetuning. Our method leverages BC to \textit{transform} heuristic-based algorithms into neural networks, enabling \textit{practical} personalization through data-efficient finetuning.

{\bf Behavioral Cloning.} In our work, we seek to imitate UKF, a rule-based system constructed from extensive domain expertise. UKF is based on an unscented Kalman filter, a common statistical algorithm used to estimate nonlinear systems. Similar to WebRTC~\cite{bergkvist2012webrtc}, UKF utilizes its filter to model the delay dynamics of the network. UKF smoothly adapts its bandwidth estimates with a set of static functions based on state variables such as the observed one-way network delays, delay gradients, and loss rates. For example, UKF incorporates a function to gradually scale its estimates in proportion to changes in one way delay.

In contrast to WebRTC's GCC, UKF was designed specifically for videoconferencing applications. Parameters determining the relative significance of loss and delay, as well as the magnitude of bandwidth estimate adjustments, were empirically tuned through extensive testing on live videoconferencing calls. Consequently, UKF's estimates do not follow an additive increase multiplicative decrease (AIMD) scheme, resulting in smoother bandwidth estimates compared to WebRTC's characteristic ``sawtooth'' behavior. UKF has been previously deployed on Microsoft Teams and shown to be competitive with alternative online algorithms (see Section~\ref{sec:sim-eval}); yet, like GCC, is difficult to tailor, making it a strong candidate to illustrate the potential of \Name{}. It is important to note that \Name{} treats reference estimators as ``black boxes'' and does not preclude the use of other heuristics, nor do we assert that UKF is the optimal bandwidth estimator. Rather, we choose to leverage UKF to illustrate how heuristic-based methods can be \textit{transformed} into flexible data-driven systems. Thus, given a set of collected heuristic demonstrations $\Xi$, network states $S$, and heuristic actions $\pi(s)$, we seek to learn a policy $\pi^*$ in the following supervised manner: 

\begin{equation}
    \pi^* = \operatorname*{arg\,min}_ {\hat{\pi}} \sum_{\xi \in \Xi} \sum_{s \in S} L(\hat{\pi}(s), \pi(s))
\end{equation}
where $\hat{\pi}$ corresponds to the policy of our imitator.

By reframing policy learning in the context of supervised learning, BC enables agents to learn a control policy while benefiting from the stability and convergence properties of supervised learning. 

\begin{figure}[t!]
    \centering
    \includegraphics[width=0.9\columnwidth, trim={0.25cm 12.4cm 0.15cm 9cm}, clip]{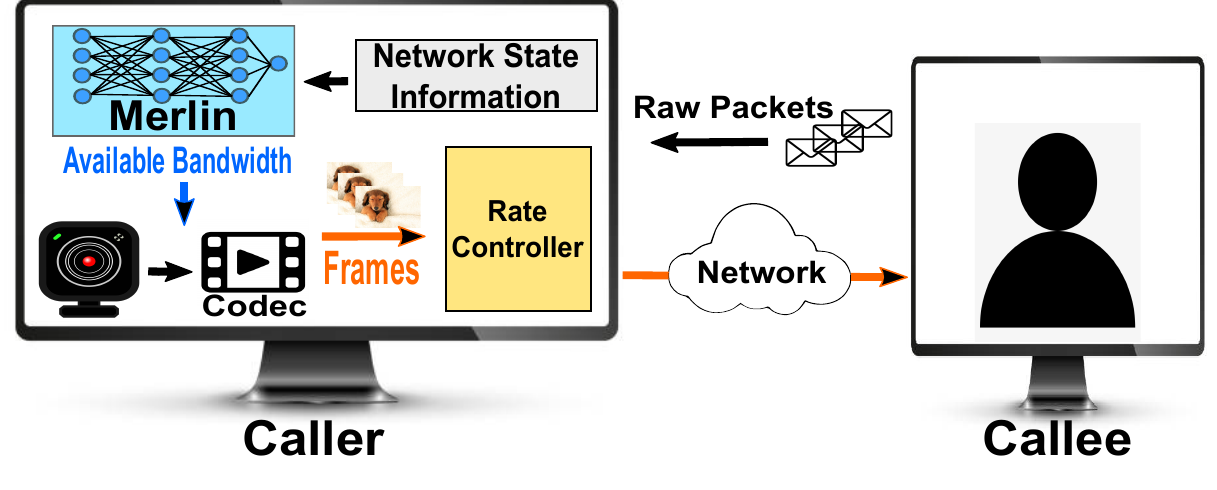}
	\caption{Media Stack Overview.}
	\label{fig:merlin_instack}
    \vspace{-6mm}
\end{figure}

{\bf Data Collection.} Despite the benefits of supervised learning, BC suffers from the problem of compounding error~\cite{bc-from-obs}. Compounding error refers to the accumulation of errors along a decision trajectory and has been shown to limit the robustness of BC models as the learned agents are incapable of bridging to new unseen environments~\cite{sage}. However, providing the imitator with a large, diverse set of demonstrations can significantly mitigate the impact of compounding error. By increasing state space coverage, the imitator is less likely to encounter unfamiliar states and accumulate errors along its trajectory. To address the limitations of BC, we collect $10000$ telemetry logs with a fork of OpenNetLab, an opensource RTC simulator~\cite{opennetlab}. Additionally, a dataset containing thousands of UKF trajectories has been released~\cite{khairy2024acm}.

Telemetry logs consisted of network state observations and UKF bandwidth estimates sampled at $60$ ms intervals. State observations were derived from aggregated packet information. Ablation studies across various feature combinations revealed that UKF features alone yielded an imitation loss of $0.007$, while the expanded feature set described below achieved a sevenfold reduction to $0.001$ (see Appendix~\ref{sec:appendix-feature-ablation} for more details). Bandwidth estimates were restricted to be between $10$ Kbps and $8$ Mbps, a range that supports audio-only calls to high-definition videoconferencing calls with screen sharing. Network simulation parameters were randomly sampled according to specific network environments, such as low bandwidth, high bandwidth, fluctuating bandwidth, burst loss, and LTE, to improve the diversity of heuristic demonstrations. Demonstrations were collected once-- prior to training.

\begin{figure}[t!]
    \centering
    \includegraphics[width=0.89\columnwidth, trim={0 10.8cm 0 6.4cm}, clip]{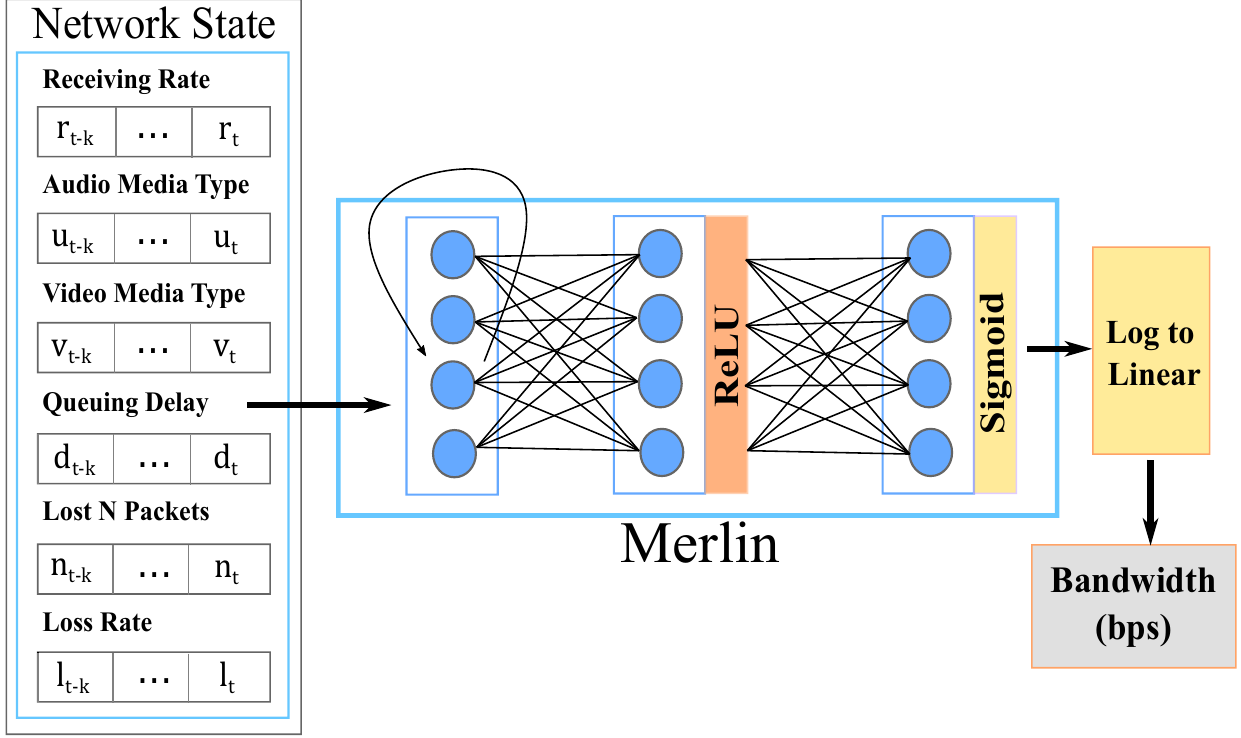}
	\caption{Merlin's Network Architecture.}
	\label{fig:arch}
    \vspace{-0.65cm}
\end{figure}

{\bf Architecture and Feature Details.} We choose to utilize a Long Short Term Memory (LSTM) for our policy network to exploit the temporal correlations present in network traffic. 
\Name{} accepts a $60$ length tensor of normalized state observations. Each state observation contains $k$ previous network measurements which serves to capture both short-term and long-term network dynamics where short-term measurements are sampled at $60$ ms granularity and long term measurements are sampled every $600$ ms. As depicted in Figure~\ref{fig:arch}, we construct state observation $s_t$ as $(\hat{r}, \hat{u}, \hat{v}, \hat{d}, \hat{n}, \hat{l})$ which corresponds to the receiving rate after applying the log transform below, proportion of audio packets, proportion of video packets, queuing delay, number of lost packets normalized by $0.05$, and the loss rate respectively. Each state variable contains $10$ measurements ($5$ short-term and $5$ long-term measurements).

\Name{} first encodes state observations with its LSTM and then leverages two fully-connected layers to decode state observations into output actions. Both the LSTM and fully-connected layers consist of $128$ neurons. The first fully-connected layer is followed by a ReLU activation function while the final output layer utilizes a sigmoid activation function, limiting the model output to a real number between $0$ and $1$. We employed an inverse log transform to project \Name{}'s output action into bps, 

\begin{equation}
    \hat{b}_{log} = \frac{\log(\hat{b}) - \log(b_{min})}{\log(b_{max}) - \log(b_{min})}
\label{eq:action_map}
\end{equation}
where $\hat{b}$ represents the estimated bandwidth, $b_{min}$ corresponds to the minimum bandwidth, $b_{max}$ is the maximum bandwidth, and all bandwidth values are in Mbps. 

{\bf Training Procedure.} 
We train \Name{} with offline UKF demonstrations. The sequence of states and actions from one call was treated as a single training sample. To train \Name{}, a batch size of $256$ was utilized, i.e., $256$ calls were ingested per training step. The mean squared error (MSE) between \Name{}'s actions and UKF's actions was minimized over the course of $1000$ training epochs. Adam optimization was employed with an initial learning rate of $0.001$. Most importantly, \Name{} never interacted with a network nor was a single packet transmitted during policy extraction.

In contrast, for online finetuning and policy personalization, we utilized $75$ randomly generated, one minute videoconferencing calls to finetune \Name{}'s extracted heuristic policy. We leveraged proximal policy optimization (PPO)~\cite{schulman_proximal_2017}, an online RL algorithm, and adopted a variant of the QoE-based reward formulation detailed in~\cite{fang_reinforcement_2019} to encode our QoE preference: 

\begin{equation}
    QoE = \alpha X + \beta R - \gamma D - \tau L
\label{eq:reward}
\end{equation}

where $X$ is the capacity utilization, $R$ is the receive rate in Mbps, $D$ is delay in seconds, and $L$ is the packet loss rate. The coefficients $\alpha$, $\beta$, $\gamma$, and $\tau$ were carefully calibrated to achieve multiple objectives. Specifically, $\alpha$ aims to optimize bandwidth efficiency while mitigating congestion, $\beta$ promotes higher receive rates to enhance audio and video quality, $\gamma$ penalizes delays to maintain interactivity, and $\tau$ discourages packet loss to ensure smoothness. This formulation was previously shown to produce an RL model that outperformed our baseline UKF in specific settings~\cite{fang_reinforcement_2019}.

{\bf Implementation.} We implement \Name{}'s LSTM architecture in Pytorch and export \Name{} to ONNX format. For live videoconferencing calls, we deploy \Name{} into our RTC media stack (see Figure~\ref{fig:merlin_instack}) for ML-based receiver-side estimation via ONNX inference. 

\section{Evaluation}
\label{sec:eval}

The key results of our evaluations are:
\begin{enumerate}
    \item \Name{} generalizes from offline telemetry logs to real-world, intercontinental videoconferencing calls, matching our heuristic's performance with no statistically significant deviations.
    \item \Name{} enables \textit{practical} personalization and improves QoE by up to $7.8$\% compared to our heuristic, while utilizing $80\%$ less training samples compared to existing online methods.
\end{enumerate}

\subsection{Methodology}

{\bf Baselines.}
We baseline against six different bandwidth estimators. (1) UKF, a hand-engineered unscented kalman filter previously deployed in production on Microsoft Teams. (2) HRCC, a hybrid bandwidth predictor that combines both RL and heuristic-based methods~\cite{hrcc}. HRCC's RL agent outputs a coefficient to tune the bandwidth estimate of its heuristic estimator. (3) Gemini~\cite{gemini}, another hybrid bandwidth predictor that incorporates an uncertainty detector to monitor variations in the observed network environment. When high network variation is detected, Gemini switches from its RL agent to a heuristic-based method. HRCC and Gemini were the winners of the ACM Multimedia Systems 2021 Bandwidth Estimation Grand Challenge. (4) BoB~\cite{bob}, a hybrid bandwidth estimator that dynamically alternates between an RL agent and a heuristic-based model. BoB was previously shown to outperform HRCC and Gemini. We provide $2$-$4$ as a reference for hybrid systems. In contrast, for our finetuning experiments, we baseline against (5) an IL model trained purely from offline experience and (6) an RL model trained online from scratch via PPO. We provide $6$ as a reference for state-of-the-art online RL algorithms~\cite{fang_reinforcement_2019}. Methods $2$-$6$ were trained in simulation; hence, to avoid introducing discrepancies between train and test conditions, we evaluate in simulation for fair comparison.

{\bf Metrics.} For our simulated evaluations, we monitor the following metrics. (1) Receiving rate: the receiving rate is reported in Mbps. (2) Packet loss rate: the percentage of lost packets. (3) Delay: the packet delay is reported in ms. (4) QoE Reward: A metric derived from congestion signals that correlates with end-user QoE~\cite{fang_reinforcement_2019}. Since calls are simulated, the simulated packets do not carry any meaningful payload; hence, we are unable to compute industry-standard QoE metrics such as the video mean opinion score (MOS)~\cite{video-mos} in this environment. Instead, we approximate user QoE with the QoE reward detailed in Equation~\ref{eq:reward}. While this QoE formulation may correlate with improved user QoE, it is important to note that the relationship between objective network metrics and subjective user QoE is ambiguous~\cite{gottipati2024balancing}. Therefore, we largely focus on QoE as a whole and provide individual network metrics as supplemental information. 

During real videoconferencing calls, we report industry-standard QoE metrics. (4) Video MOS: the MOS values range from $1$ to $5$ with $1$ indicating poor QoE and $5$ indicating exceptional QoE. We utilize a vision-based model to produce a video MOS estimate based on the received video signals. The estimates were shown to exhibit $99\%$ correlation with user visual experience based on the ITU-T's P.910 scores~\cite{video-mos}. (5) Audio MOS: similar to the video MOS, the audio MOS also ranges from $1$ to $5$ with $1$ indicating low audio QoE and $5$ indicating high QoE. We utilize a learned audio model to produce an audio MOS estimate based on the received audio signals. The estimates were internally shown to correlate significantly with user audio experience based on ITU-T's P.808 scores. We refer to video MOS and audio MOS as video QoE and audio QoE respectively.

\subsection{Imitating the Heuristic}
\label{sec:sim-eval}

We measure the performance of \Name{} against UKF over real networks in the wild. Our setup consists of $20$ nodes distributed across North America, Asia, and Europe. For each evaluation call, we randomly sample $10$ pairs of nodes. We then conduct calls with UKF 
and \Name{}. Our experiments were executed during the day and at night over the course of a week. We conducted hundreds of evaluation runs and utilized a Welch t-test to determine whether our results were statistically significant. Our evaluation consists of over $700$ in the wild calls per model ($\approx 1400$ calls in total). Our key metrics with their $95\%$ confidence interval are summarized in Table~\ref{tab:planetstat}.

\begin{table}[t!]
\centering
\resizebox{\columnwidth}{!}{%
\begin{tabular}{@{}lrrr@{}}
\toprule
\textbf{Metric} & \multicolumn{1}{c}{\textbf{Merlin}} & \multicolumn{1}{c}{\textbf{UKF}} & \multicolumn{1}{c}{\textbf{p-value}} \\
\midrule
Video QoE & 4.06 $\pm$ 0.62 & 4.19 $\pm$ 0.32 & \textbf{0.97} \\
Audio QoE & 4.81 $\pm$ 0.13 & 4.82 $\pm$ 0.12 & \textbf{0.99} \\
Receiving Rate (Mbps) & 1.91 $\pm$ 0.70 & 2.15 $\pm$ 0.31 & \textbf{0.95} \\
Delay Mean (ms) & 68.5 $\pm$ 26.72 & 71.9 $\pm$ 15.51 & \textbf{0.68} \\
Loss Rate (\%) & 0.0021 $\pm$ 0.0136 & 0.0025 $\pm$ 0.0177 & \textbf{0.83} \\
\bottomrule
\end{tabular}%
}
\caption{A/B Comparison of Merlin and UKF over Intercontinental Videoconferencing Calls}
\label{tab:planetstat}
\vspace{-6mm}
\end{table}

\begin{figure*}[t!]
\centering
\begin{subfigure}[t]{0.24\textwidth}
\centering
\includegraphics[width=\textwidth, trim={0 0cm 0 0cm}, clip]{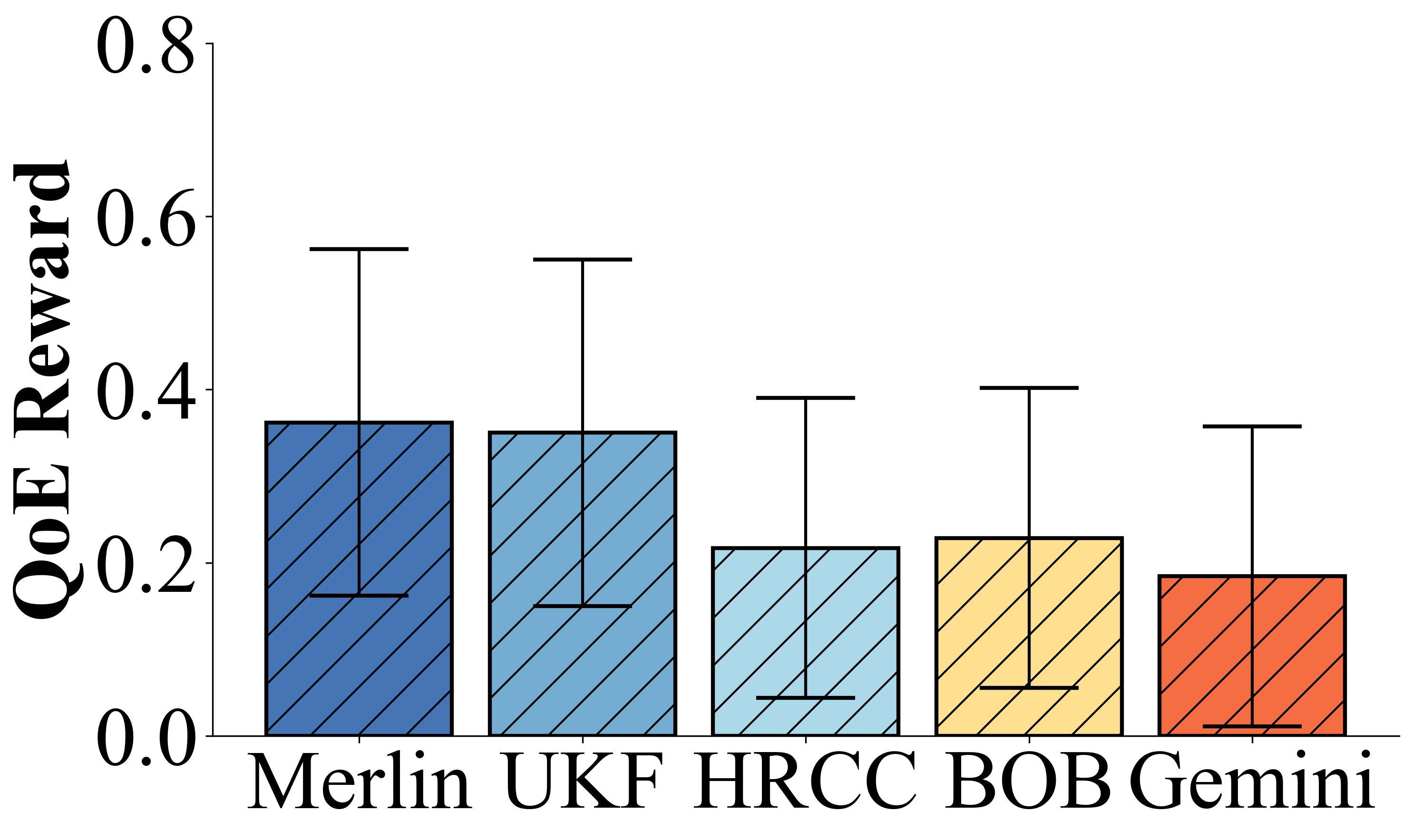}
\caption{QoE Reward}
\label{fig:qoe_reward}
\end{subfigure}
\begin{subfigure}[t]{0.24\textwidth}
\centering
\includegraphics[width=\textwidth, trim={0 0cm 0 0cm}, clip]{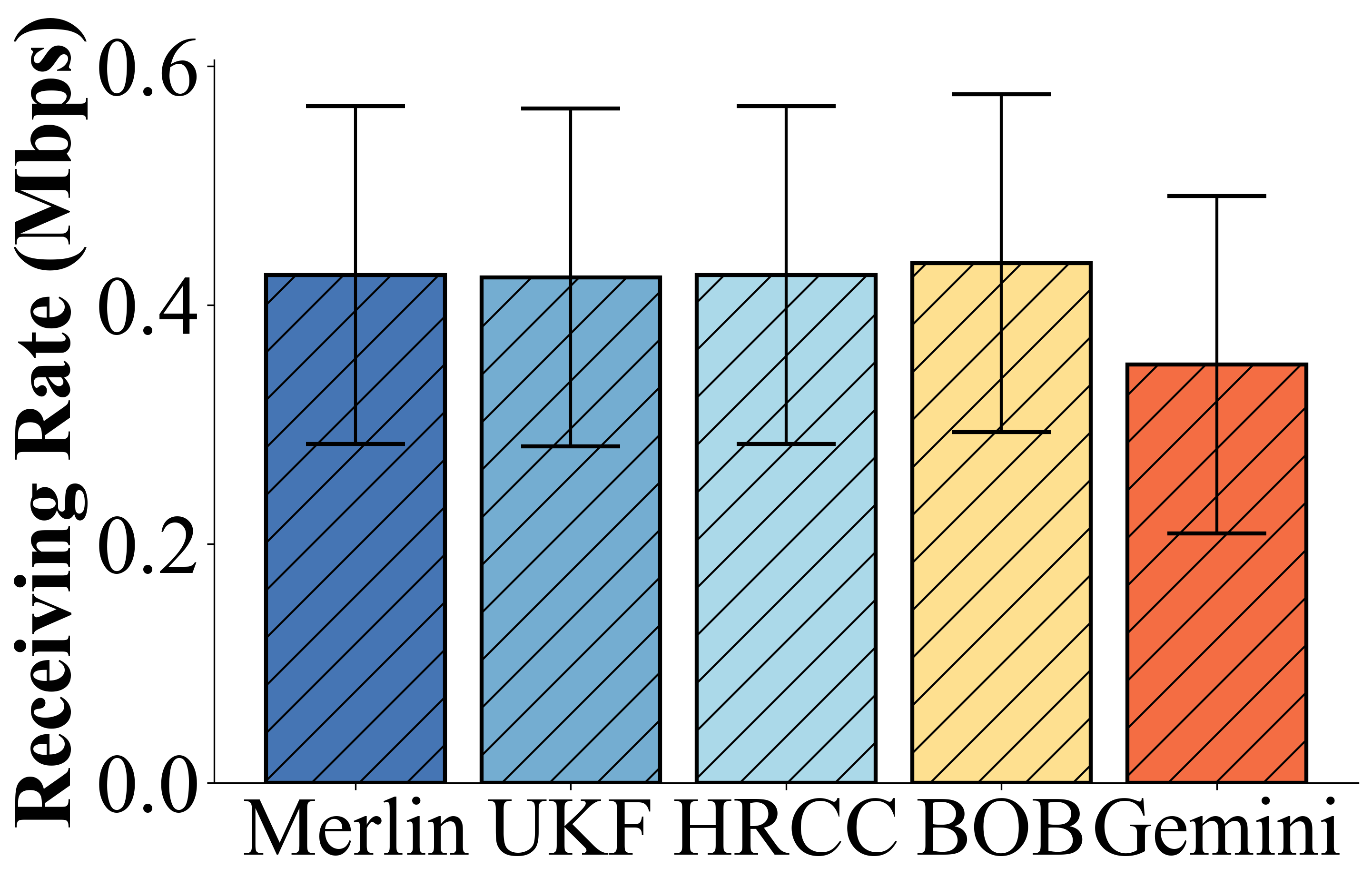}
\caption{Receiving Rate}
\label{fig:receive_rate_sim}
\end{subfigure}
\hfill
\begin{subfigure}[t]{0.24\textwidth}
\centering
\includegraphics[width=\textwidth, trim={0 0cm 0 0cm}, clip]{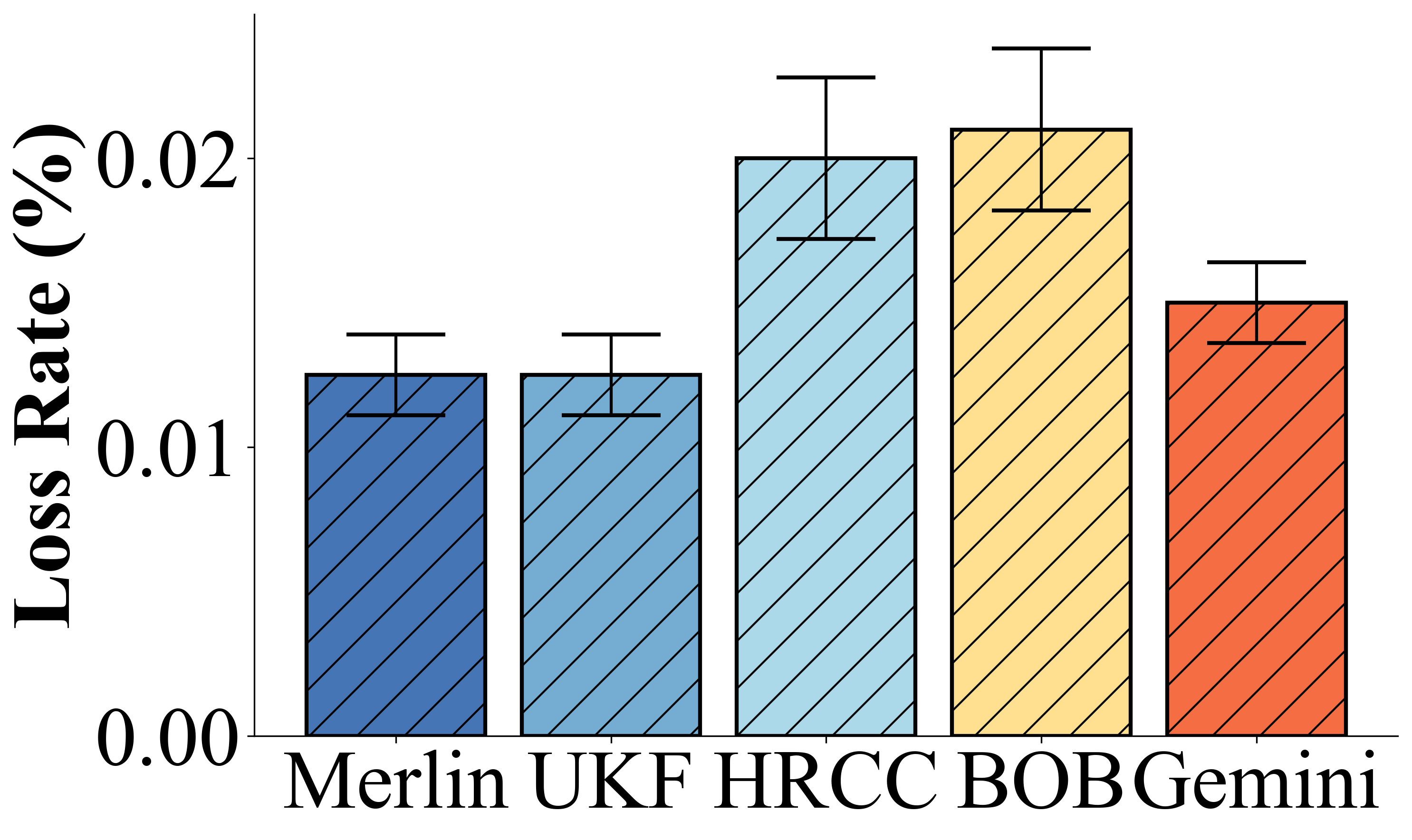}
\caption{Loss Rate}
\label{fig:packet_loss_sim}
\end{subfigure}
\hfill
\begin{subfigure}[t]{0.24\textwidth}
\centering
\includegraphics[width=\textwidth, trim={0 0cm 0 0cm}, clip]{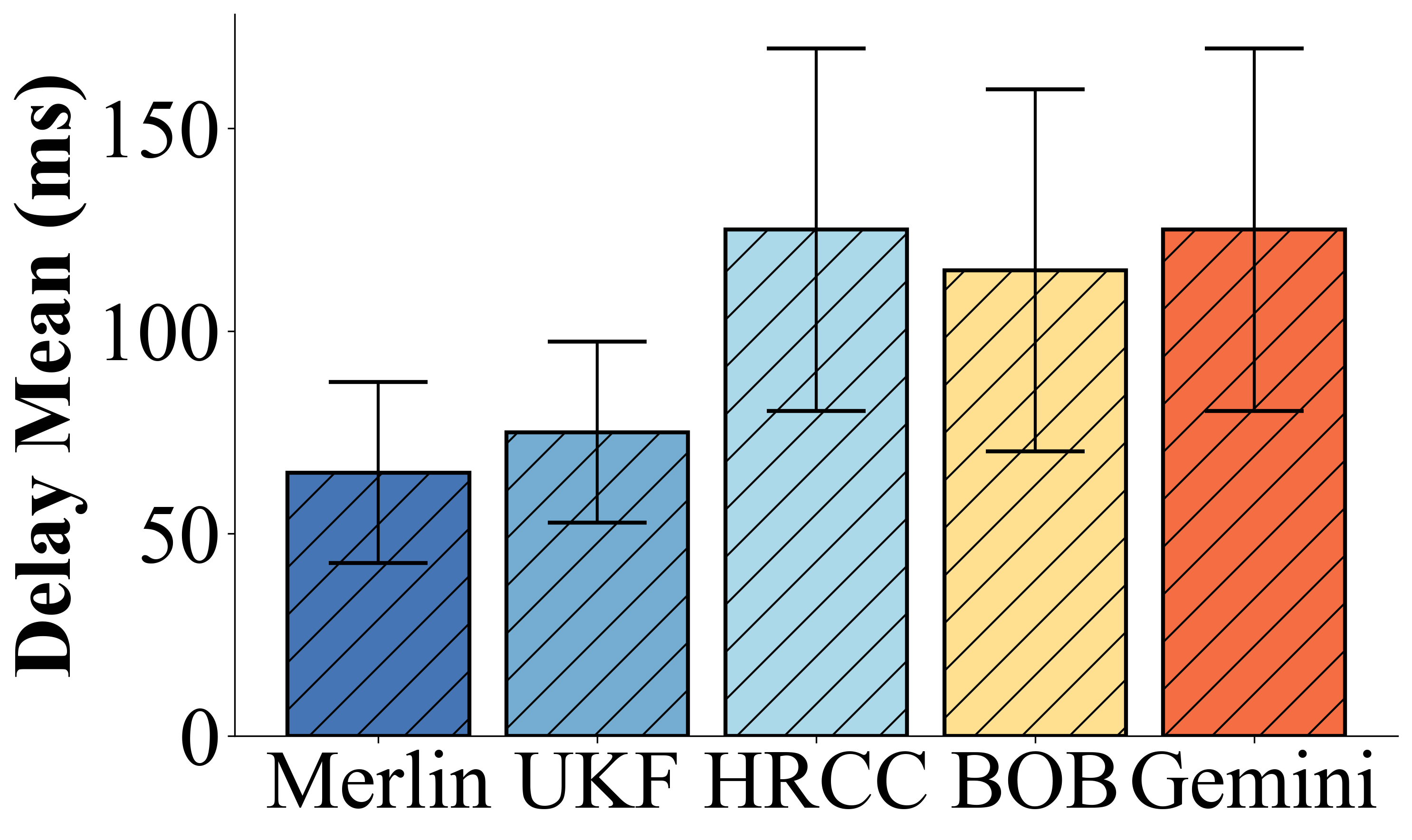}
\caption{Delay Mean}
\label{fig:delay_sim}
\end{subfigure}
\caption{Performance Comparison of \Name{}, UKF, and Hybrid Methods during Videoconferencing Calls.}
\vspace{-5mm}
\label{fig:sim_baselines}
\end{figure*}

The results demonstrate that \Name{} is capable of generalizing from offline experience to real-world environments, performing inline with our heuristic UKF. Additionally, we observe no statistically significant difference in video QoE and audio QoE (i.e., p-values exceed $0.05$ in Table~\ref{tab:planetstat}) between \Name{} and UKF during real intercontinental videoconferencing calls. This indicates that \Name{} is retaining the heuristic policy even within environments that differ significantly from our offline experience. Despite our heuristic not being observed in intercontinental environments, we hypothesize that the combination of carefully selected state features and a diverse demonstration set enables \Name{} to effectively transition from offline telemetry logs to in-the-wild deployments.

We further verify our method by comparing against HRCC, Gemini, and BoB using traces generated from production parameters within a fork of OpenNetLab~\cite{opennetlab}. We randomly generate traces for evaluation from a diverse set of network workloads containing low bandwidth, high bandwidth, fluctuating bandwidth, burst loss, and LTE. Call simulation parameters such as the proportion of video calls to audio calls, network queuing delay, and the video start time are randomly sampled at runtime. Our evaluation consists of $480$ distinct simulated calls. We run $1000$ validation runs, which corresponds to nearly $480000$ simulated calls.

Our results are reported in Figure~\ref{fig:sim_baselines}. No significant difference in the receiving rate
was observed between bandwidth estimators; however, both \Name{} and UKF achieve a significant reduction in the packet loss rate and delay in comparison to BoB, HRCC, and Gemini. Specifically, we observe approximately $44\%$ less delay on average and a $17\%$ reduction in the packet loss rate, which translates to a $58\%$ improvement in QoE compared to competing methods (see Figure~\ref{fig:sim_baselines}). Our experiments indicate no statistically significant difference between \Name{} and UKF and demonstrate that \Name{} generalizes from offline data to online videoconferencing calls-- 
with zero network interactions during training. Furthermore, our results highlight UKF's expertise as a bandwidth estimator.

\subsection{Personalizing Merlin}
\label{sec:pretrain}
\begin{figure*}[t]
\centering
\begin{subfigure}[b]{0.28\textwidth}
\centering
\includegraphics[width=0.9\textwidth, trim={0 0cm 0 0cm}, clip]{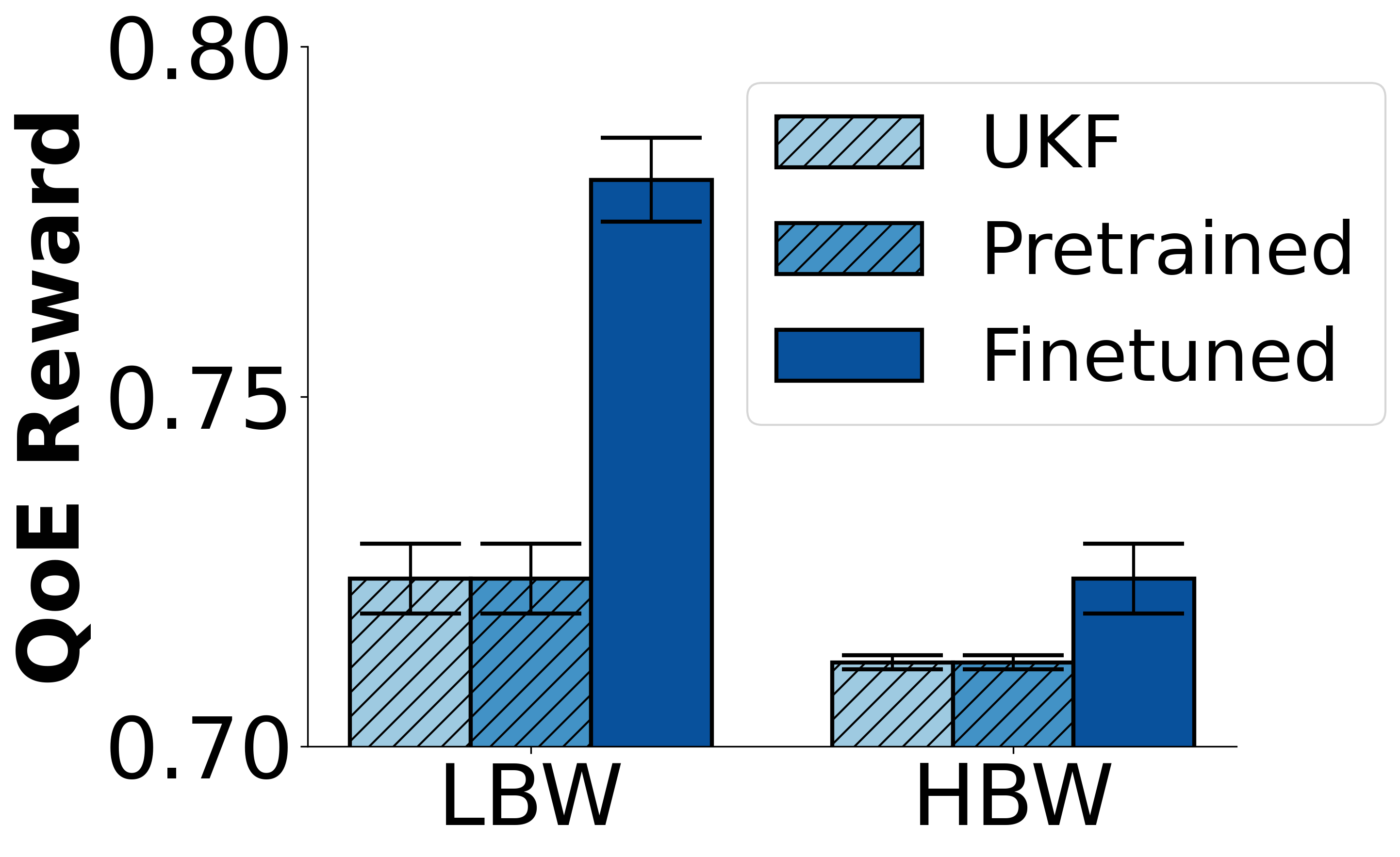}
\caption{QoE Reward.}
\label{fig:ukf-merlin-qoe}
\end{subfigure}
\hfill
\begin{subfigure}[b]{0.35\textwidth}
\centering
\includegraphics[width=0.9\textwidth, trim={0 0cm 0 0cm}, clip]{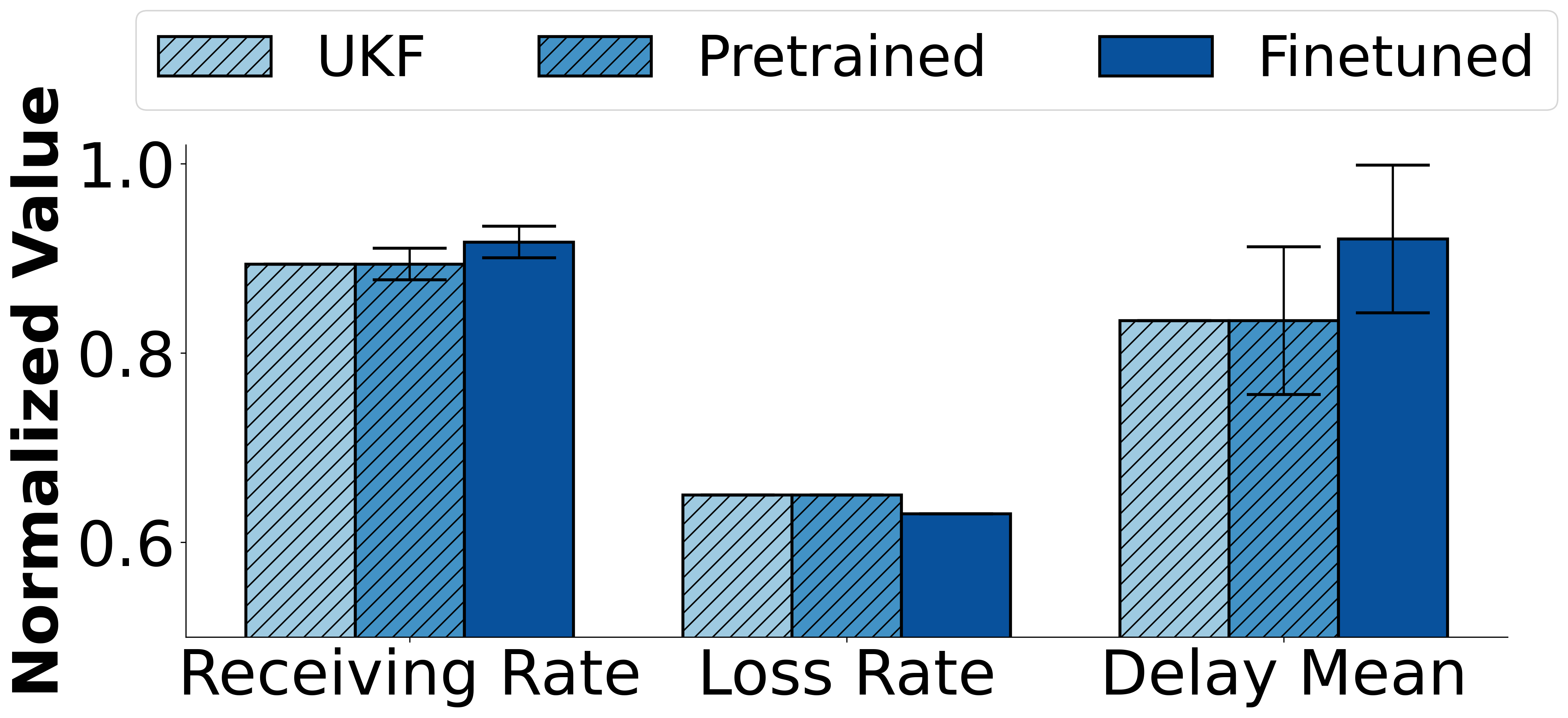}
\caption{HBW Network Metrics.}
\label{fig:ukf-merlin-network-hbw}
\end{subfigure}
\hfill
\begin{subfigure}[b]{0.35\textwidth}
\centering
\includegraphics[width=0.9\textwidth, trim={0 0cm 0 0cm}, clip]{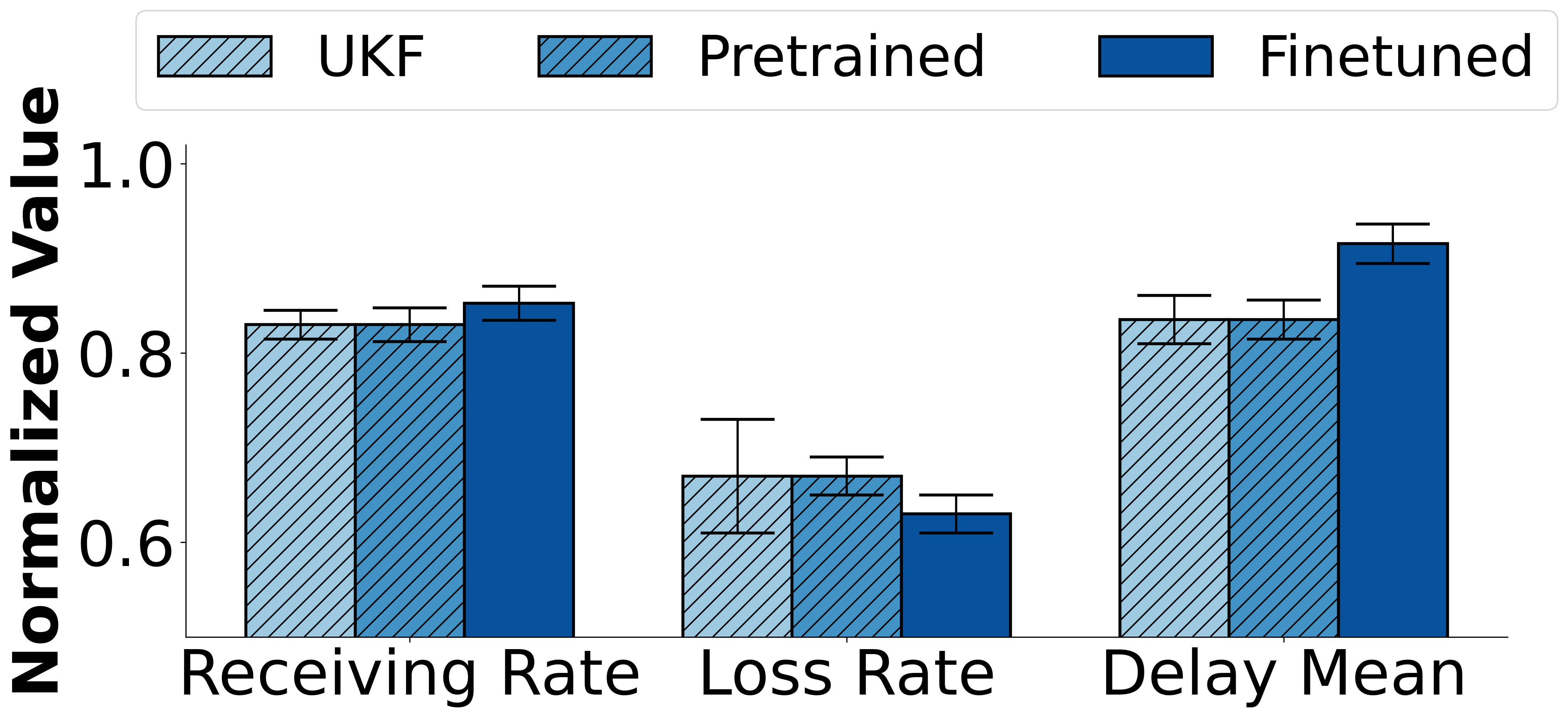}
\caption{LBW Network Metrics.}
\label{fig:ukf-merline-network-lbw}
\end{subfigure}
\caption{Personalizing the Heuristic Policy to LBW and HBW environments.}
\label{fig:fine-tune}
\vspace{-5mm}
\end{figure*}

To illustrate the potential of \Name{}, we hypothesize the promise for improving user QoE by tailoring our heuristic policy to end user environments. 
We target two sets of end user environment distributions: low bandwidth (LBW) with less than $1$ Mbps capacity and high bandwidth (HBW) with at least $1$ Mbps capacity. That is, we personalize \Name{} to each environment through online finetuning. We leverage a total of $75$ online calls drawn from either LBW or HBW environments for finetuning. The learned estimators are evaluated on holdout sets drawn from their respective target distribution: a set of $30$ LBW traces or a set of $30$ HBW traces. Both target distributions encompass stable, burst loss, and fluctuating bandwidth scenarios. We vary the seed and repeat the experiment five times. Our results are summarized in Figure~\ref{fig:fine-tune}. We now refer to the model trained solely from offline experience as the \textit{pretrained} model and refer to the model tuned via online experience as the \textit{finetuned} model.

Quantitatively, in both LBW and HBW scenarios, the \textit{finetuned} model achieves a higher receiving rate in exchange for mixed results in terms of loss and delay in comparison to UKF. This specialization results in a $7.8\%$ and $1.4\%$ improvement in QoE over our heuristic UKF in LBW and HBW environments, respectively (Figure~\ref{fig:ukf-merlin-qoe}). 
This exercise illustrates the following key aspects of \Name{}. One, in contrast to learning a policy from scratch, \Name{} learns a bandwidth estimation policy by pretraining on offline telemetry logs. Two, \Name{}'s policy can then be finetuned to match a desired QoE preference purely from new network observations, improving end-user QoE. This workflow stands in stark contrast to handcrafted methods such as UKF which require extensive domain knowledge to tune the carefully calibrated parameters. By combining the flexibility of data-driven methods with the domain knowledge of traditional heuristics, \Name{}'s IL-based design facilitates \textit{practical} personalization.

\subsection{Offline to Online vs Online}

\begin{figure*}[t]
\centering
\begin{subfigure}[b]{0.33\textwidth}
\centering
\includegraphics[width=0.9\textwidth, trim={0 0cm 0 0cm}, clip]{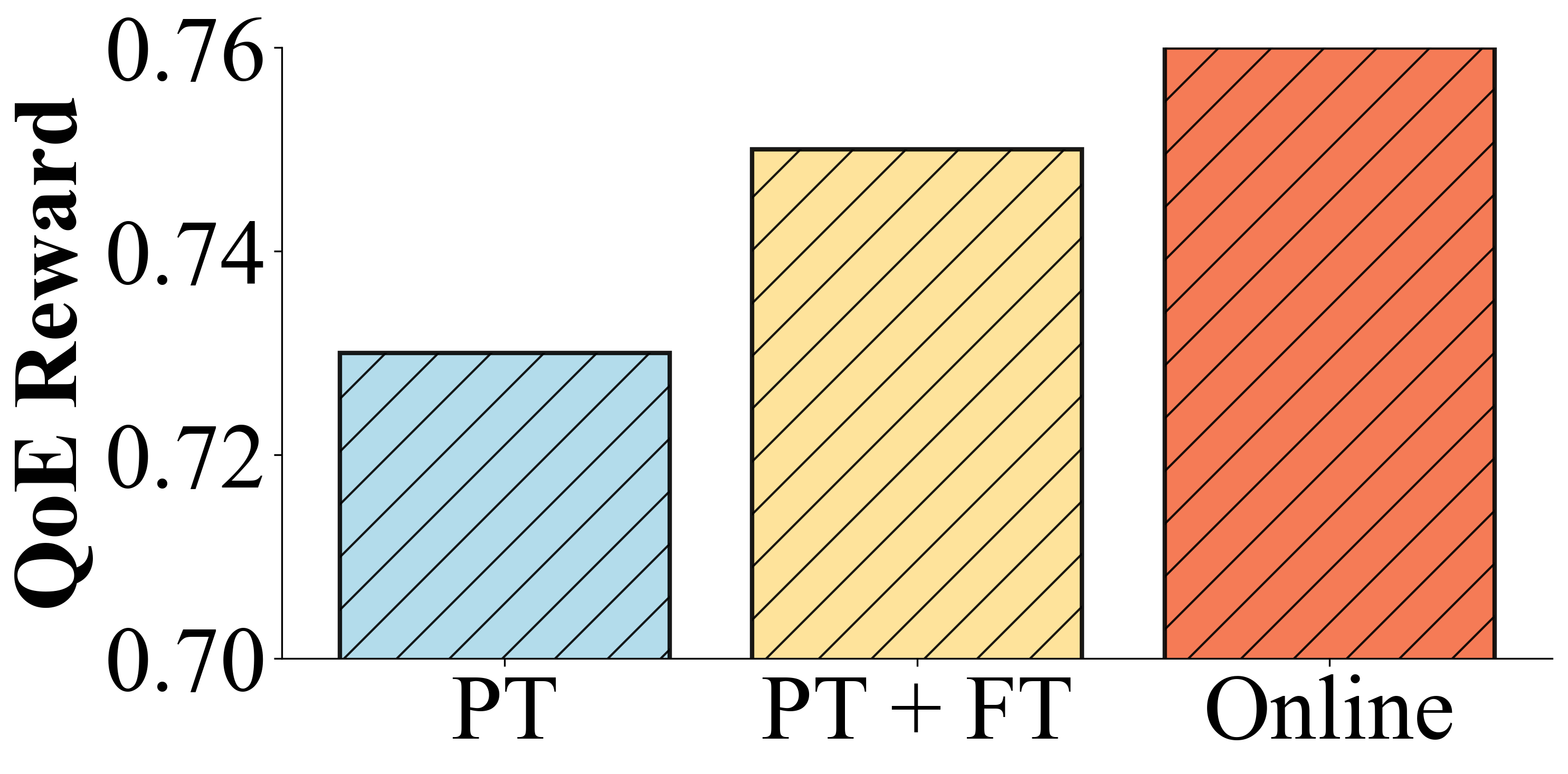}
\caption{Max QoE Reward.}
\label{fig:max-qoe-reward}
\end{subfigure}
\hfill
\begin{subfigure}[b]{0.33\textwidth}
\centering
\includegraphics[width=0.9\textwidth, trim={0 0cm 0 0cm}, clip]{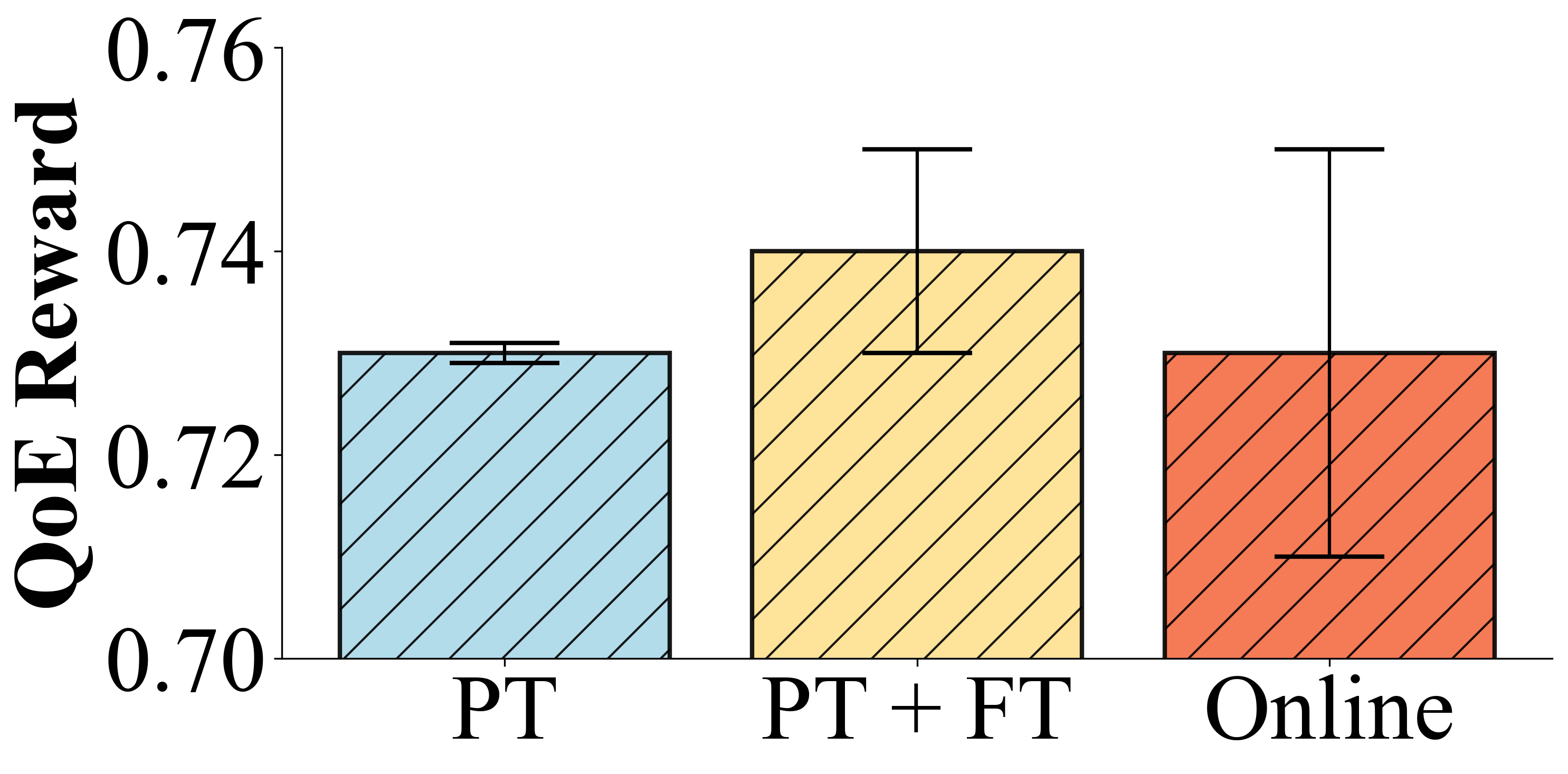}
\caption{Mean QoE Reward.}
\label{fig:mean-qoe-reward}
\end{subfigure}
\hfill
\begin{subfigure}[b]{0.3\textwidth}
\centering
\includegraphics[width=0.9\textwidth, trim={0 0cm 0 0cm}, clip]{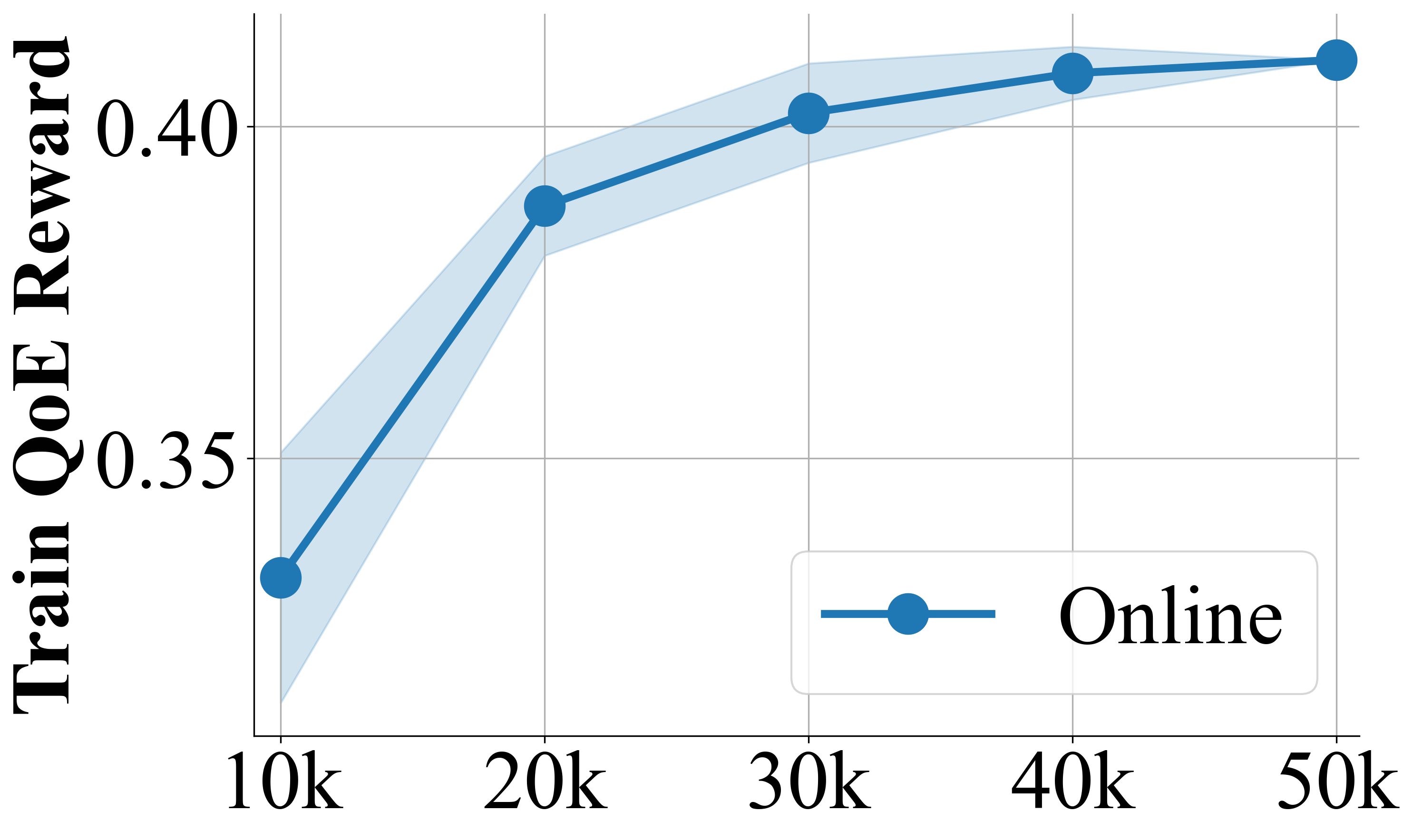}
\caption{QoE vs. Videoconferencing Calls.}
\label{fig:train-online-reward}
\end{subfigure}
\caption{Pretraining + Finetuning (PT + FT) vs. Online Training.}
\label{fig:online-finetune}
\vspace{-5mm}
\end{figure*}

We provide a comparison between online training and offline pretraining with online finetuning to illustrate the practical benefits of our approach. We refer to these models as {\it online} and {\it finetuned} models respectively. We train $5$ variants of each model, varying the initial KL penalty of PPO from $\{0.1, 0.3, 0.5, 0.7, 0.9\}$ where a lower KL penalty encourages exploration and allows for larger policy updates. The learned estimators are evaluated on a holdout set of $60$ LBW and HBW traces that encompass stable, burst loss, and fluctuating bandwidth scenarios. We report the aggregated results in Figure~\ref{fig:online-finetune}. It is important to note that we utilized $10000$ calls for pretraining and $75$ calls for finetuning the \textit{finetuned} models, while the \textit{online} models leveraged the same model architecture as \Name{} and were trained until convergence via PPO.

We empirically observe higher variance in QoE early on in the training procedure with lower KL penalties due to increased exploration (Figure~\ref{fig:train-online-reward}). Our results indicate that the \textit{finetuned} models achieve a competitive maximum QoE and a higher mean QoE in comparison to the best \textit{online} models, which suggests that the \textit{finetuned} policies are of similar quality (Figure~\ref{fig:mean-qoe-reward}). We hypothesize that pretraining provides a rich prior which enables the agent to converge to a competitive set of policies.

In addition to quality, we emphasize that, in our case, the time-to-solution through pretraining and finetuning is significantly lower than online training. All \textit{online} models continue to improve with more call samples and converge after approximately $50000$ videoconferencing calls on average as the variance in training QoE reward approaches $0$ (Figure~\ref{fig:train-online-reward}). In contrast, our \textit{finetuned} models converge with $10075$ sampled calls ($10000$ for pretraining and $75$ for finetuning), a near $80\%$ reduction in sample complexity. Online RL algorithms such as PPO are considered the gold standard due to their ability to directly learn from the target environment; however, because they begin from a blank slate, they require a substantial number of samples to converge~\cite{schulman_proximal_2017}. Conversely, our method, which employs offline IL with finetuning and relies largely on supervised learning, demonstrates convergence with a significantly reduced sample size. Our results suggest that offline pretraining with finetuning leads to a quality set of policies with significant reductions in the time-to-solution compared to online training from scratch. Thus, we can achieve competitive performance with state-of-the-art online RL but with an $80\%$ reduction in sample complexity.


\section{Conclusion}
\label{sec:conclusion}

Our work tackles key challenges in adopting AI-based systems for real-time network control. We propose
\Name{}, a novel solution to BWE that \textit{transforms} heuristic-based methods into data-driven neural networks, enabling \textit{practical} personalization through finetuning. Through a series of experiments, we illustrate that \Name{} generalizes from offline telemetry logs to the real world, showing that prior domain-rich control policies can be distilled into deep learning-based models via offline IL. We then provide results demonstrating the promise of personalizing heuristic-based network control policies through finetuning as opposed to manual engineering. Future research on data curation can help \Name{} learn niche network characteristics and improve end user QoE by selectively finetuning on the most informative network observations.
\section*{Acknowledgements}
We would like to thank Lili Qiu for her highly valuable feedback. We thank Scott Inglis and Ezra Ameri for their help.
\bibliographystyle{IEEEtran}
\bibliography{IC3-AI, bibliography}

\appendix

\section{Ablation Studies}
\label{sec:appendix-feature-ablation}

\begin{figure}[t!]
\centering
\includegraphics[width=0.9\columnwidth, trim={0 0.25cm 0 0.3cm}, clip]{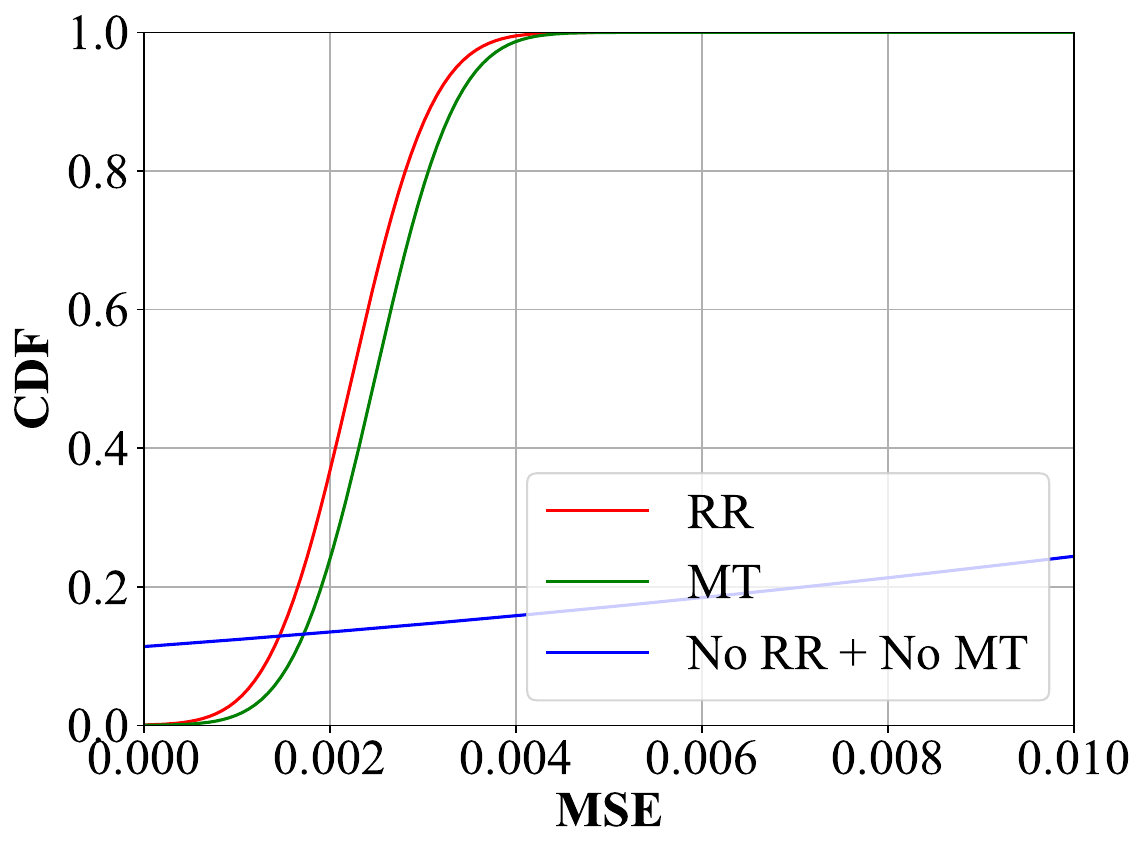}
\vspace{-2mm}
\caption{Imitation Performance with Different Feature Sets: Receive Rate subsets (RR); Media Type subsets (MT); No Receive Rate or Media Type features (No RR + No MT).}
\vspace{-4mm}
\label{fig:feat-cdf}
\end{figure}

{\bf Impact of Features.} We seek to study the impact of different features on learned BWE. We group features into five categories and ablate on these groups: receiving rate, loss ratio, average number of lost packets, queuing delay, and media type features. The media type features report the probability mass of video packets and audio packets over the last five time steps. We then exhaustively retrain \Name{} on each feature subset and report the performance on generated traces within our simulated environment in Figure~\ref{fig:feat-cdf}. Our experiments indicate that the two most impactful features are the receiving rate and the media type features. Surprisingly, we find that both of these features alone are sufficient to mimic UKF. We find that the best performing subset of features contains all five feature groups. We emphasize that the best performing feature subset is distinct from the handcrafted features delivered to UKF; hence, simply reusing the expert congestion signals for training may be suboptimal.

\begin{figure}[h!]
\centering
\begin{subfigure}[h!]{\columnwidth}
\centering
\includegraphics[width=0.8\columnwidth, trim={0 0.15cm 0 0}, clip]{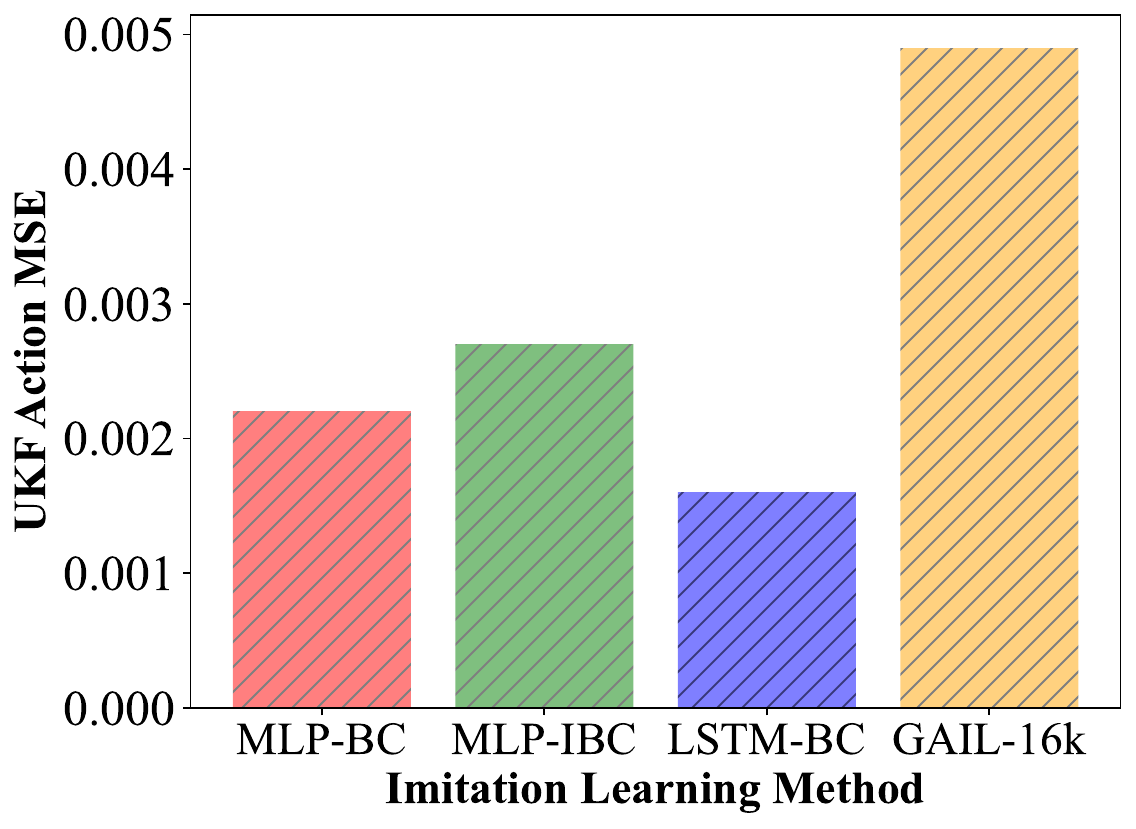}
\caption{IL Methods.}
\label{fig:il-method-gym}
\end{subfigure}
\hfill
\begin{subfigure}[h!]{\columnwidth}
\centering
\includegraphics[width=0.8\columnwidth, trim={0 0.1cm 0 0}, clip]{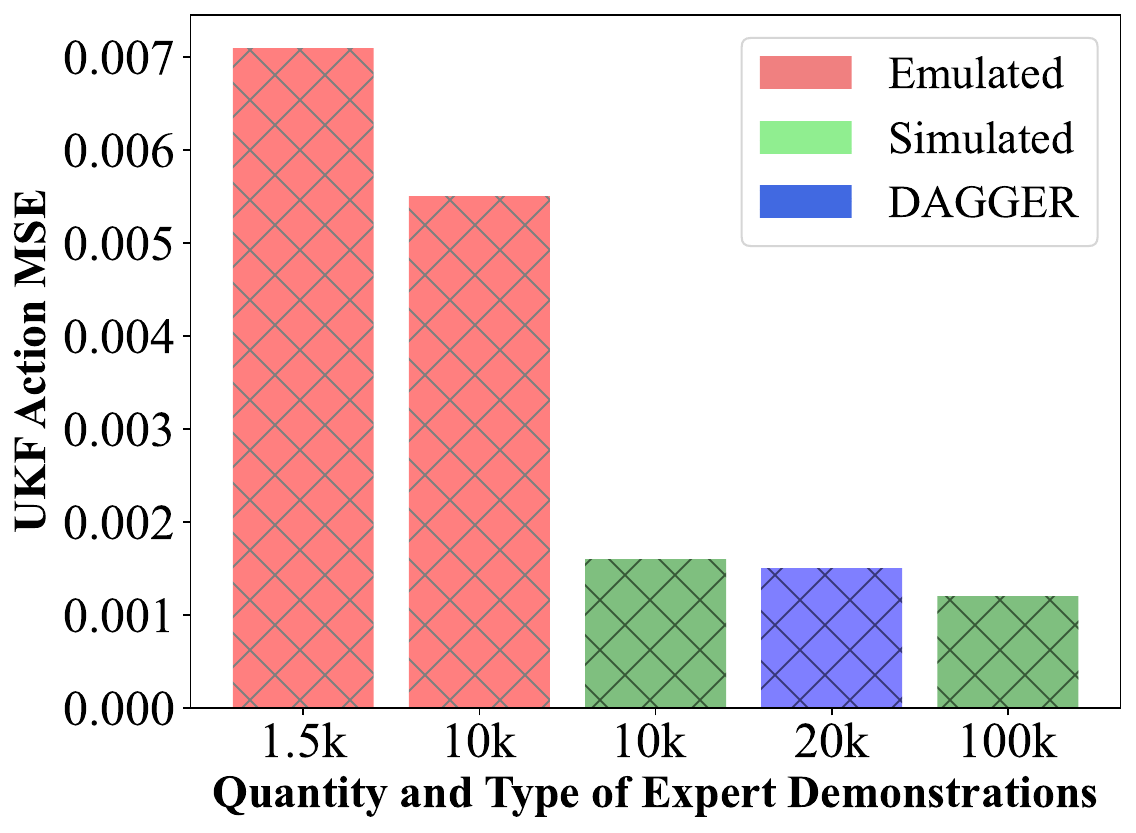}
\caption{Data Quantity and Quality.}
\label{fig:data-size}
\end{subfigure}
\caption{IL Ablation Study Results.}
\label{fig:all-ablation}
\end{figure}

{\bf Learning Methods and Architecture.} We evaluate the performance of different IL methods and architectures for BWE within our gym environment. We compare three different IL approaches: BC, Implicit Behavioral Cloning (IBC)~\cite{ibc}, and Generative Adversarial Imitation Learning (GAIL)~\cite{gail} and two policy architectures for BC: a multi-layer perceptron (MLP) and an LSTM. We report our findings in Figure~\ref{fig:il-method-gym}. We implement IBC and GAIL with MLP-based policy networks. We adopt similar hyperparameters for each model and maintain the same gym validation parameters across trials. We train GAIL with $16000$ expert state-action samples per training epoch. Our evaluations indicate that BC with an LSTM policy network outperforms all other benchmarked methods (see Figure~\ref{fig:il-method-gym}). 

{\bf Quantity and Quality of Demonstrations.}
We explore the impact of the quantity and quality of demonstrations on our learned estimator. Specifically, we compare with UKF demonstrations drawn from two environments: our simulation and an emulated environment. We retrain \Name{} on four different datasets: $1500$ emulated trajectories, $10000$ emulated trajectories, $10000$ simulated trajectories, and $100000$ simulated trajectories. The emulated trajectories are intentionally limited; that is, we do not randomize call parameters and leverage parameters drawn directly from production traces. By limiting the breadth of demonstrations from the target distribution, the relation between demonstration diversity and data quality can be explored. Emulated data is of higher quality as the domain shift between offline and online samples is reduced.

Furthermore, we evaluate the implications of DAGGER~\cite{dagger} on our $10000$ gym dataset, doubling the dataset size to $20000$ demonstrations by the end of the training run. We report our findings in Figure~\ref{fig:data-size}. We observe that the model trained with $100000$ demonstrations performs the best overall. While we observe that $10000$ demonstrations are sufficient to learn a BWE policy, increasing the number of heuristic observations appears to improve generalization. Furthermore, data diversity appears to impact imitation performance more than demonstration quality, i.e., heuristic demonstrations should be collected from a wide variety of workloads and network environments. Thus, to learn network control policies directly from offline telemetry logs, we find that providing the agent with a large number of diverse demonstrations is key to ensuring robustness against domain shift which is corroborated in~\cite{sage}. 

\end{document}